\documentclass{PoS}

\def\MS{\hbox{$\overline{\rm MS}$}}

\def\QMS{Q$_0$\MS}
\def\PR{{\it Phys.~Rev.~}}

\def\NP{{\it Nucl.~Phys.~}}

\def\PL{{\it Phys.~Lett.~}}

\def\SJNP{{\it Sov.~Jour.~Nucl.~Phys.~}}

\def\JHEP{{\it Jour.~High~Energy~Phys.~}}
\def\vol#1{{\bf #1}}\def\vyp#1#2#3{\vol{#1} (#2) #3}

\catcode`@=11 
\def\slash#1{\mathord{\mathpalette\c@ncel#1}}
 \def\c@ncel#1#2{\ooalign{$\hfil#1\mkern1mu/\hfil$\crcr$#1#2$}}
\def\lsim{\mathrel{\mathpalette\@versim<}}
\def\gsim{\mathrel{\mathpalette\@versim>}}
 \def\@versim#1#2{\lower0.2ex\vbox{\baselineskip\z@skip\lineskip\z@skip
       \lineskiplimit\z@\ialign{$\m@th#1\hfil##$\crcr#2\crcr\sim\crcr}}}

\title{Structure Function Resummation in small-x QCD}

\ShortTitle{Small-x QCD}

\author{\speaker{Guido Altarelli}\\
 Dipartimento di Fisica ``E.Amaldi'' and  
Sezione INFN, Universit\`a Roma Tre,  Roma, Italy\\
and CERN, Department of Physics, Theory Division, Gen\`eve, Switzerland\\
E-mail: \email{guido.altarelli@cern.ch}}

\author{Richard D. Ball\\
School of Physics, University of Edinburgh,
Edinburgh EH9 3JZ, Scotland\\
and CERN, Department of Physics, Theory Division, Gen\`eve, Switzerland\\
E-mail: \email{rdb@ph.ed.ac.uk}}

\author{Stefano Forte\\
Dipartimento di Fisica e Sezione INFN, Universit\`a di
Milano, Milan, Italy\\
E-mail: \email{stefano.forte@mi.infn.it}}

\abstract{We summarize our recent results on small $x$ resummation
in full QCD with $n_f$ quark flavours and discuss their 
phenomenological impact in the extraction of
parton distributions from present day structure function data and their
extrapolation to the kinematics relevant for future colliders such as
the LHC.}

\FullConference{8th International Symposium on Radiative Corrections (RADCOR)\\
		 October 1-5 2007\\
		 Florence, Italy}

\begin{document}

\noindent 
Higher order calculations in perturbative QCD have progressed in an 
extraordinary way
in recent years, motivated  by the needs of accurate phenomenology
at the LHC. The frontier of
present-day perturbative calculations is the next-to-next-to leading
(NNLO) order~\cite{gehrrev}, due to the availability of a variety
of novel computational techniques. Results for cross-sections 
at NNLO can be used
thanks to the recent
determination~\cite{nnlo} of three--loop splitting
functions which drive NNLO perturbative evolution. 
However, perturbative evolution at NNLO is
unstable in the high energy (small $x$ limit): the size of the NNLO
corrections diverges as $x\rightarrow 0$ at fixed scale. 

Small $x$ resummation, which should take care of this instability, has
a rather long history, starting with the original determination of
leading high energy corrections~\cite{bfkl} and their inclusion in
perturbative  anomalous dimensions~\cite{jar}. Until quite recently, 
however, its
relevance for phenomenology has been modest: since the advent of HERA
data, it is clear that a NLO description of observed scaling violation
is perfectly adequate~\cite{das,heralhc}, and the data show no
evidence for large small $x$ effects.  The reason why nominally large
corrections seem to have no impact has been obscure for a long
time. However, due to some accidental zeros in
coefficients,  small $x$ contributions to
NLO perturbative evolution are small, so in practice
one could simply ignore the issue for all practical 
purposes. 
At  NNLO, however, small $x$ terms are large, the perturbative
instability is manifest, and resummation becomes mandatory.

Fortunately,
over the last few years a fully resummed approach to
perturbative evolution has been
constructed. Within this approach, it is possible
to understand why   fixed
perturbative order 
corrections are very large at small
$x$, yet their full resummation leads to a considerable softening of small
$x$ terms, consistent with the fact that the HERA data do not show 
any large departure from NLO predictions.  In
order to obtain stable resummed results one must
satisfy various
physical constraints, such as momentum conservation, renormalization
group invariance and gluon exchange symmetry. These require the
inclusion of several classes
of terms which are formally subleading in comparison to the series of
leading or next-to-leading small $x$ logs. Once these constraints are
enforced, the resummed perturbative expansion at small $x$ becomes
stable, and one no longer sees the strong small $x$ enhancement or
suppression that the leading~\cite{bfkl} and subleading~\cite{fl}-\cite{dd} 
small $x$ logs would give.

A comparison of existing approaches
to this resummation, as discussed respectively in
refs.~\cite{sxap,sxres,sxrun} (ABF approach)
and~\cite{salam,ciaf} (CCSS approach),
shows~\cite{heralhc} that they
yield results which agree with each other within the expected
theoretical uncertainty. They agree in including the physical
assumptions listed above. They
differ mostly because the CCSS approach is built up within the BFKL
framework, by improving the BFKL kernel through the inclusion of terms
which become important in the collinear region,
while the ABF approach is based on the construction of an improvement
of the GLAP anomalous dimension through the inclusion of terms that
become important in the small $x$ region. The fact that they lead to
very similar results is thus a consequence of the ``duality'' which
relates the BFKL and GLAP description of perturbative evolution: at
leading twist, they can describe
the same physics, provided the respective evolution kernels
are suitably matched~\cite{sxrun,afp,rcdual}\footnote{Small-$x$ corrections to
polarized structure functions, of different form and origin than  those
considered here, have been recently discussed in ref.~\cite{pdl}}.

Because the leading high energy corrections are dominated by gluon
exchange, the resummation is most easily performed in the pure Yang-Mills
theory, and indeed the full resummed results for perturbative
evolution of refs. \cite{sxap, ciaf} have been obtained with
$n_f=0$. In order to actually get predictions for (flavour singlet) physical
observables, one needs  the full two-by-two matrix of splitting
functions, and the set of hard coefficient functions for the desired
observables, in a given factorization scheme. The construction of
resummed splitting functions is simplified by the fact that only one
of the two eigenvalues of the singlet anomalous dimension matrix has
leading $N=0$ singularities: hence, only this eigenvalue is affected
by the resummation. In deep--inelastic scattering, the construction
of resummed coefficient functions is further simplified by the fact
that virtual photons at leading order only couple to quarks. This
implies that there always exist schemes where only one parton (quark
or gluon) contributes to each of the structure functions $F_2$ and
$F_L$. It follows that in any factorization scheme the
resummation of the coefficient is specified in terms of a single
function for each structure function (these functions have been
determined in ref.~\cite{ch}, where they are called
$h_2$ and $h_L$ for $F_2$ and $F_L$ respectively.)

Hence, (at least) two strategies are available for the construction of
resummed observables in deep-inelastic scattering. The first
possibility is to simply pick a factorization scheme, then
construct resummed two-by-two evolution kernels  and 
resummed coefficient functions in that scheme. This program was started in
ref.~\cite{matevol}, where 
the full  $n_f\not=0$ resummed evolution matrix was constructed
by extending a BFKL--like approach to coupled quark and gluon
evolution, along the lines of the  
approach of refs. \cite{salam, ciaf}. 
This has the advantage of giving evolution equations for
off-shell, unintegrated parton distributions, but it has the
shortcoming of providing results in a factorization scheme which 
only coincides with \MS\ up to the
next-to-leading fixed order, and differs from it at the resummed
level. Available resummed coefficient functions~\cite{ch}, which are given in
\MS\ or DIS,  are not readily combined with the evolution kernels
determined in this way. 
Similar problems were encountered in ref.~\cite{thorne}, where
resummed structure functions were obtained by combining resummed
anomalous dimensions and coefficients determined in different
factorizaton schemes.

\begin{figure}
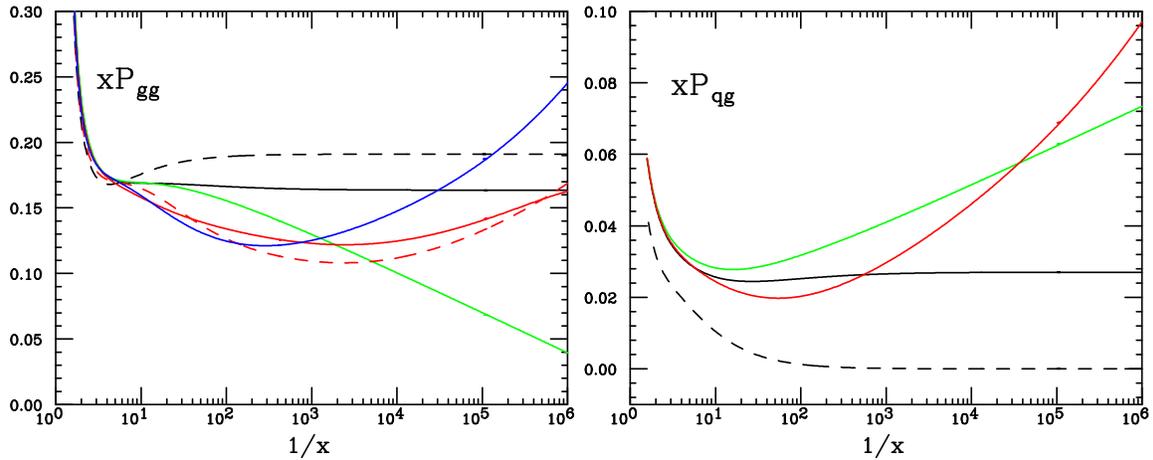
 
\includegraphics[width=.5 \textwidth]{pgg.ps} 
\includegraphics[width=.5 \textwidth]{pqg.ps} 
\caption{The splitting functions $xP_{gg}$ and $xP_{qg}$ for $n_f=4$ and 
$\alpha_s=0.2$ as a 
function of $x$.  Fixed order perturbation theory LO (black, dashed);
NLO (black, solid) and NNLO (green, solid)  and resummed LO (red, dashed)
and NLO in \QMS\ scheme (red, solid) and in the 
\MS\ scheme (blue, solid).} 
\label{fig1} 
\end{figure} 

\begin{figure}
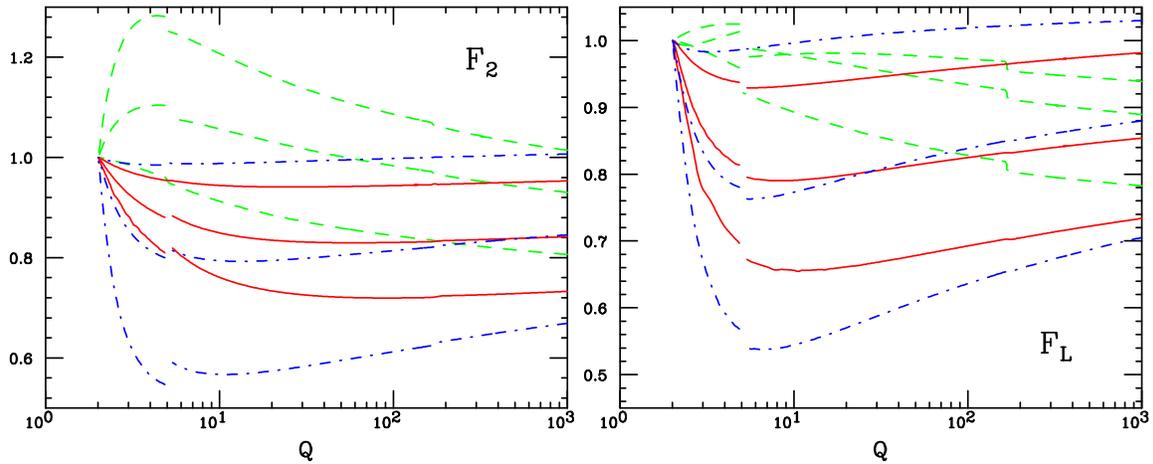
 
\includegraphics[width=.5 \textwidth]{f2.ps} 
\includegraphics[width=.5 \textwidth]{fl.ps} 
\caption{The 
$K$-factors singlet $F_2$  and $F_L$ structure functions at 
fixed  $x=10^{-2},~10^{-4}~{\rm or}~10^{-6}$ as a
function of $Q$ with $\alpha_s$ running 
and $n_f$ varied (the breaks in the curves correspond to the b and 
t quark thresholds). Fixed order perturbation theory NNLO (green, dashed);
resummed NLO  in \QMS\ scheme (red, solid),
resummed NLO in the \MS\ scheme (blue, dot-dashed).} 
\label{fig2} 
\end{figure} 
A second possibility
consists of taking advantage of the peculiar fact that both the
resummation of the anomalous dimensions and that of the
deep--inelastic coefficients for a given structure function are each
determined in terms of a single function: it is only when specifying
the factorization scheme that a matrix of anomalous dimensions and a
vector of coefficients for each structure function are obtained. It
follows that if one has full control of the transformation between
factorization schemes at the resummed level, one can obtain results
for resummed coefficient functions and anomalous dimensions in any
scheme. 

This approach was pursued in ref.~\cite{symphen}, where by making 
use of the general techniques for cross-section resummation 
developed in ref.~\cite{ballhad} we were able to 
compute the full $n_f\not=0$  matrix of resummed 
singlet splitting functions (see fig.~1) 
together with the  coefficients $C^i_q$ and 
$C^i_g$ (i=2,L) for the singlet structure functions $F_2$ and $F_L$, 
respectively, in various commonly used schemes such as the \MS\
scheme. 
This then allowed us to determine for the first time
the structure functions at a fully resummed level in a consistent
factorization scheme. While referring to ref.~\cite{symphen} for a
full discussion of the resummed splitting functions and coefficient
functions, we focus here on the impact of the resummation on physical
observables, and on the extraction of parton distributions from them.

\begin{figure}
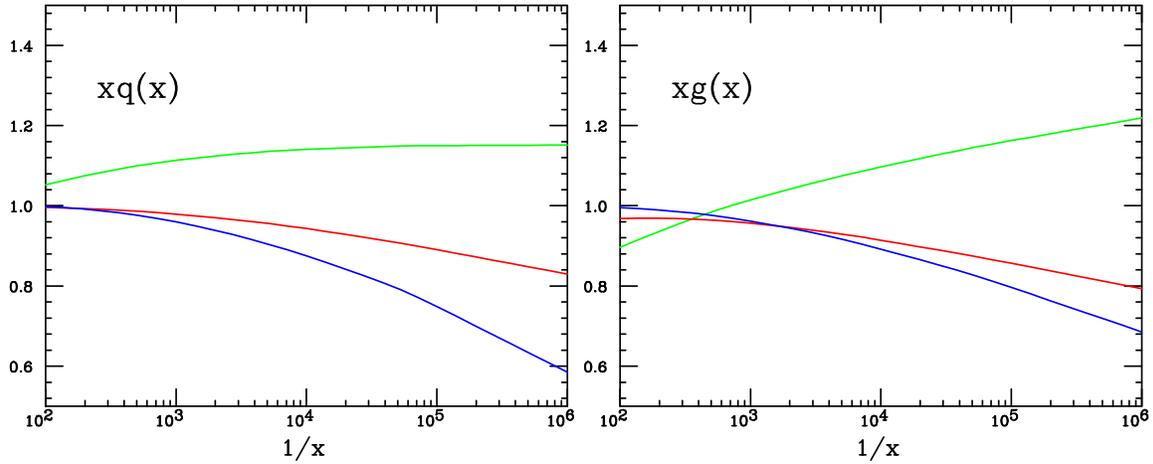
 
\includegraphics[width=.5 \textwidth]{qbcx.ps} 
\includegraphics[width=.5 \textwidth]{gbcx.ps} 
\caption{
The 
$K$-factors, defined as the ratio of the
fixed order NNLO or resummed to the NLO fixed order result,
 for  $xq$ and 
$xg$ when $F_2$ and $F_L$ are fixed at the  reference scale
$Q_0=5$~GeV. Results are shown at $Q=Q_0$ as 
as function of $x$ in the range $x=10^{-2}~{\rm or}~10^{-6}$. The
curves (top to bottom at small $x$) are:
fixed order perturbation theory NNLO (green);
resummed NLO  in \QMS\ scheme (red),
resummed NLO in the \MS\ scheme (blue).
} 
\label{fig3} 
\end{figure} 
Thus, in fig.~2 we  compare the effects 
of resummed evolution with those of 
fixed order perturbative evolution at NLO (which is the usual baseline 
for most current fits) and also at NNLO. We show the $K$-factor 
(defined as resummed/(NLO fixed order pert)) for the singlet
 $F_2$  structure function at fixed values of $x$ as a 
function of $Q$. For each $x$ value 
we present three curves: the resummed case in the \QMS\ scheme, 
the corresponding plot in the \MS\ scheme, and the NNLO fixed order 
perturbative. The starting point at $Q_0=2$ is taken to be the same 
for the structure functions plotted in the three curves. 
In the expanded linear scale of these 
plots we can see the moderate scheme difference between the 
\QMS\ and the \MS\ schemes, which is left after combining 
coefficients and parton densities, which, individually, show a 
much more pronounced scheme dependence.

For $F_2$ the effect of resummation, at sufficiently small 
$x$ values, goes in the opposite direction to the NNLO perturbative 
evolution: the resummed $K$-factor is less than one, corresponding to a 
smaller structure function at higher scales than with fixed order 
perturbative NLO evolution. The effect of the resummation is somewhat
larger than that of the NNLO. This shows that for $x\lsim 0.2$ the
inclusion of unresummed NNLO terms is actually counterproductive.

In view of using parton distributions extracted from HERA data for
physics at the LHC, it is also interesting to study how the quark and
gluon distributions change when the resummation is switched on, while
imposing that the measurable structure functions be unchanged. We do
this in fig.~3, where we display the $K$-factors for the input quark and
gluon distributions as a function of $x$, with the input scale 
$Q_0$=5~GeV. Here the
structure functions $F_2$ and $F_L$ at the input scale are kept fixed,
but the coefficient functions are changed depending on the 
perturbative approximation employed.  Whereas
at NNLO the quark and gluon distributions are enhanced with respect to
the NLO, at the resummed level they are suppressed; the suppression
becomes increasingly significant at smaller $x$. The effect on the
parton distribution is sizable, but when evolving up the
scale dependence tends to reduce this effect leading to the more
moderate corrections displayed in figure~2. The general conclusion,
however, is that if resummation effects are disregarded, the
associated error in extracting parton distributions at HERA and
evolving them up at LHC is of order of about 5\% at $x\sim 10^{-3}$,
and as large as 20\% for low values of $x\sim10^{-6}$.

\bigskip
\noindent{\bf Acknowledgements}

Two of us, G.A. and S.F., acknowledge that the present work has been 
supported in part by the Italian Ministero dell'Universita' e della 
Ricerca Scientifica, under the PRIN program for 2007-08. The work of R.D.B.  
has been done in the context of the Scottish Universities' Physics Alliance.
This work was partly supported by the Marie Curie Research and Training 
network HEPTOOLS under contract MRTN-CT-2006-035505.

\end{document}